\pdfoutput=1

\documentclass[12pt,a4paper]{article}

\usepackage{ifthen} 
\newboolean{pdflatex}
\setboolean{pdflatex}{true} 

\newboolean{articletitles}
\setboolean{articletitles}{true} 

\newboolean{uprightparticles}
\setboolean{uprightparticles}{false} 

\newboolean{inbibliography}
\setboolean{inbibliography}{false} 

\newboolean{wordcount}
\setboolean{wordcount}{false} 

\ifthenelse{\boolean{wordcount}}{ 

\usepackage{xspace} 
\usepackage{upgreek}

\def\lhcb {\mbox{LHCb}\xspace}

\def\MagUp {\mbox{\em Mag\kern -0.05em Up}\xspace}

\ifthenelse{\boolean{uprightparticles}}%
{

 \def\Ppi         {\ensuremath{\uppi}\xspace}

 \def\PDelta      {\ensuremath{\Delta}\xspace}                 
 \def\PXi      {\ensuremath{\Xi}\xspace}                 
 \def\PLambda      {\ensuremath{\Lambda}\xspace}                 
 \def\PSigma      {\ensuremath{\Sigma}\xspace}                 
 \def\POmega      {\ensuremath{\Omega}\xspace}                 
 \def\PUpsilon      {\ensuremath{\Upsilon}\xspace}                 
 
 \def\PB      {\ensuremath{\mathrm{B}}\xspace}                 
                  
 \def\PD      {\ensuremath{\mathrm{D}}\xspace}

 \def\PK      {\ensuremath{\mathrm{K}}\xspace}

 \def\Pb      {\ensuremath{\mathrm{b}}\xspace}                 
 \def\Pc      {\ensuremath{\mathrm{c}}\xspace}

 \def\Pi      {\ensuremath{\mathrm{i}}\xspace}

 \def\Pp      {\ensuremath{\mathrm{p}}\xspace}                 
 \def\Pq      {\ensuremath{\mathrm{q}}\xspace}                 
                  
 \def\Ps      {\ensuremath{\mathrm{s}}\xspace}                 
                  
 \def\Pu      {\ensuremath{\mathrm{u}}\xspace}

}
{

 \def\Ppi         {\ensuremath{\pi}\xspace}

 \mathchardef\PDelta="7101
 \mathchardef\PXi="7104
 \mathchardef\PLambda="7103
 \mathchardef\PSigma="7106
 \mathchardef\POmega="710A
 \mathchardef\PUpsilon="7107
                  
 \def\PB      {\ensuremath{B}\xspace}                 
                  
 \def\PD      {\ensuremath{D}\xspace}

 \def\PK      {\ensuremath{K}\xspace}

 \def\Pb      {\ensuremath{b}\xspace}                 
 \def\Pc      {\ensuremath{c}\xspace}

 \def\Pi      {\ensuremath{i}\xspace}

 \def\Pp      {\ensuremath{p}\xspace}                 
 \def\Pq      {\ensuremath{q}\xspace}                 
                  
 \def\Ps      {\ensuremath{s}\xspace}                 
                  
 \def\Pu      {\ensuremath{u}\xspace}

}

\makeatletter
\ifcase \@ptsize \relax
  \newcommand{\miniscule}{\@setfontsize\miniscule{4}{5}}
\or
  \newcommand{\miniscule}{\@setfontsize\miniscule{5}{6}}
\or
  \newcommand{\miniscule}{\@setfontsize\miniscule{5}{6}}
\fi
\makeatother

\DeclareRobustCommand{\optbar}[1]{\shortstack{{\miniscule (\rule[.5ex]{1.25em}{.18mm})}
  \\ [-.7ex] $#1$}}

\def\quark     {{\ensuremath{\Pq}}\xspace}

\def\uquark    {{\ensuremath{\Pu}}\xspace}

\def\squark    {{\ensuremath{\Ps}}\xspace}

\def\cquark    {{\ensuremath{\Pc}}\xspace}

\def\bquark    {{\ensuremath{\Pb}}\xspace}

\def\pion   {{\ensuremath{\Ppi}}\xspace}

\def\pip    {{\ensuremath{\pion^+}}\xspace}
\def\pim    {{\ensuremath{\pion^-}}\xspace}

\def\kaon    {{\ensuremath{\PK}}\xspace}
  \def\Kbar    {{\kern 0.2em\overline{\kern -0.2em \PK}{}}\xspace}

\def\KorKbar    {\kern 0.18em\optbar{\kern -0.18em K}{}\xspace}

\def\Kp      {{\ensuremath{\kaon^+}}\xspace}
\def\Km      {{\ensuremath{\kaon^-}}\xspace}

\def\Kstarzb {{\ensuremath{\Kbar{}^{*0}}}\xspace}

  \def\Dbar    {{\kern 0.2em\overline{\kern -0.2em \PD}{}}\xspace}

\def\DorDbar    {\kern 0.18em\optbar{\kern -0.18em D}{}\xspace}

\def\Bbar    {{\ensuremath{\kern 0.18em\overline{\kern -0.18em \PB}{}}}\xspace}

\def\BorBbar    {\kern 0.18em\optbar{\kern -0.18em B}{}\xspace}

  \def\Y#1S{\ensuremath{\PUpsilon{(#1S)}}\xspace}

\def\proton      {{\ensuremath{\Pp}}\xspace}

\def\Xires       {{\ensuremath{\PXi}}\xspace}
\def\Xiresbar    {{\ensuremath{\overline \Xires}}\xspace}
\def\Lz          {{\ensuremath{\PLambda}}\xspace}
\def\Lbar        {{\ensuremath{\kern 0.1em\overline{\kern -0.1em\PLambda}}}\xspace}
\def\LorLbar    {\kern 0.18em\optbar{\kern -0.18em \PLambda}{}\xspace}

\def\Sigmares    {{\ensuremath{\PSigma}}\xspace}

\def\Omegares    {{\ensuremath{\POmega}}\xspace}

\def\Lc      {{\ensuremath{\Lz^+_\cquark}}\xspace}

\def\Xic     {{\ensuremath{\Xires_\cquark}}\xspace}

\def\Xicp    {{\ensuremath{\Xires^+_\cquark}}\xspace}

\def\Xicbarm {{\ensuremath{\Xiresbar{}_\cquark^-}}\xspace}
\def\Omegac    {{\ensuremath{\Omegares^0_\cquark}}\xspace}

\def\to                 {\ensuremath{\rightarrow}\xspace}

\def\AT#1     {\ensuremath{A_{\mathrm{T}}^{#1}}\xspace}           

\def\C#1      {\ensuremath{\mathcal{C}_{#1}}\xspace}                       
\def\Cp#1     {\ensuremath{\mathcal{C}_{#1}^{'}}\xspace}                    
\def\Ceff#1   {\ensuremath{\mathcal{C}_{#1}^{\mathrm{(eff)}}}\xspace}        
\def\Cpeff#1  {\ensuremath{\mathcal{C}_{#1}^{'\mathrm{(eff)}}}\xspace}       
\def\Ope#1    {\ensuremath{\mathcal{O}_{#1}}\xspace}                       
\def\Opep#1   {\ensuremath{\mathcal{O}_{#1}^{'}}\xspace}                    

\newcommand{\tev}{\ifthenelse{\boolean{inbibliography}}{\ensuremath{~T\kern -0.05em eV}}{\ensuremath{\mathrm{\,Te\kern -0.1em V}}}\xspace}
\newcommand{\gev}{\ensuremath{\mathrm{\,Ge\kern -0.1em V}}\xspace}
\newcommand{\mev}{\ensuremath{\mathrm{\,Me\kern -0.1em V}}\xspace}
\newcommand{\kev}{\ensuremath{\mathrm{\,ke\kern -0.1em V}}\xspace}
\newcommand{\ev}{\ensuremath{\mathrm{\,e\kern -0.1em V}}\xspace}
\newcommand{\gevc}{\ensuremath{{\mathrm{\,Ge\kern -0.1em V\!/}c}}\xspace}
\newcommand{\mevc}{\ensuremath{{\mathrm{\,Me\kern -0.1em V\!/}c}}\xspace}
\newcommand{\gevcc}{\ensuremath{{\mathrm{\,Ge\kern -0.1em V\!/}c^2}}\xspace}
\newcommand{\gevgevcccc}{\ensuremath{{\mathrm{\,Ge\kern -0.1em V^2\!/}c^4}}\xspace}
\newcommand{\mevcc}{\ensuremath{{\mathrm{\,Me\kern -0.1em V\!/}c^2}}\xspace}

\def\invfb   {\ensuremath{\mbox{\,fb}^{-1}}\xspace}

\newcommand{\chisq}{\ensuremath{\chi^2}\xspace}

\newcommand{\chisqip}{\ensuremath{\chi^2_{\mathrm{IP}}}\xspace}

\def\gsim{{~\raise.15em\hbox{$>$}\kern-.85em
          \lower.35em\hbox{$\sim$}~}\xspace}
\def\lsim{{~\raise.15em\hbox{$<$}\kern-.85em
          \lower.35em\hbox{$\sim$}~}\xspace}

\def\PDF {PDF\xspace}

\def\pt         {\mbox{$p_{\mathrm{ T}}$}\xspace}

\def\tell1  {TELL1\xspace}
\def\ukl1   {UKL1\xspace}

\newcommand{\ie}{\mbox{\itshape i.e.}\xspace}

  \def\pkpi     {\ensuremath{\proton \Km \pip}\xspace}
  \newcommand{\al}{\ensuremath{\kern 0.5em }}
  \newcommand{\all}{\ensuremath{\kern 0.25em }}
  \usepackage{bibentry} 
  \usepackage{mciteplus} 
  \usepackage{comment} 
  \excludecomment{align*} 
  \excludecomment{align} 
  \excludecomment{equation*} 
  \excludecomment{equation} 
  \excludecomment{eqnarray*} 
  \excludecomment{eqnarray} 
}{
}
\ifthenelse{\boolean{wordcount}}{}{

\usepackage[top=1in, bottom=1.25in, left=1in, right=1in]{geometry}

\columnsep=5mm
\addtolength{\belowcaptionskip}{0.5em}

\raggedbottom
\sloppy

\usepackage{microtype}
\usepackage{lineno}  
\usepackage{xspace} 
\usepackage{caption} 

\usepackage{graphicx}  
\usepackage{color}
\usepackage{colortbl}
\graphicspath{{./figs/}} 

\usepackage{amsmath} 
\usepackage{amssymb}
\usepackage{amsfonts}
\usepackage{upgreek} 

\newcommand*\patchAmsMathEnvironmentForLineno[1]{%
\expandafter\let\csname old#1\expandafter\endcsname\csname #1\endcsname
\expandafter\let\csname oldend#1\expandafter\endcsname\csname
end#1\endcsname
 \renewenvironment{#1}%
   {\linenomath\csname old#1\endcsname}%
   {\csname oldend#1\endcsname\endlinenomath}%
}
\newcommand*\patchBothAmsMathEnvironmentsForLineno[1]{%
  \patchAmsMathEnvironmentForLineno{#1}%
  \patchAmsMathEnvironmentForLineno{#1*}%
}
\AtBeginDocument{%
\patchBothAmsMathEnvironmentsForLineno{equation}%
\patchBothAmsMathEnvironmentsForLineno{align}%
\patchBothAmsMathEnvironmentsForLineno{flalign}%
\patchBothAmsMathEnvironmentsForLineno{alignat}%
\patchBothAmsMathEnvironmentsForLineno{gather}%
\patchBothAmsMathEnvironmentsForLineno{multline}%
\patchBothAmsMathEnvironmentsForLineno{eqnarray}%
}

\usepackage{hyperref}    
\usepackage[all]{hypcap} 

\def\pkpi     {\ensuremath{\proton \Km \pip}\xspace}
\def\calL         {{\ensuremath{\cal L}\xspace}}
\newcommand{\al}{\ensuremath{\kern 0.5em }}
\newcommand{\all}{\ensuremath{\kern 0.25em }}
\usepackage{cite} 
\usepackage{mciteplus}

}
\usepackage{longtable} 

\begin{document}

\renewcommand{\thefootnote}{\fnsymbol{footnote}}
\setcounter{footnote}{1}

\ifthenelse{\boolean{wordcount}}{}{

\begin{titlepage}
\pagenumbering{roman}

\vspace*{-1.5cm}
\centerline{\large EUROPEAN ORGANIZATION FOR NUCLEAR RESEARCH (CERN)}
\vspace*{1.5cm}
\noindent
\begin{tabular*}{\linewidth}{lc@{\extracolsep{\fill}}r@{\extracolsep{0pt}}}
\ifthenelse{\boolean{pdflatex}}
{\vspace*{-2.7cm}\mbox{\!\!\!\includegraphics[width=.14\textwidth]{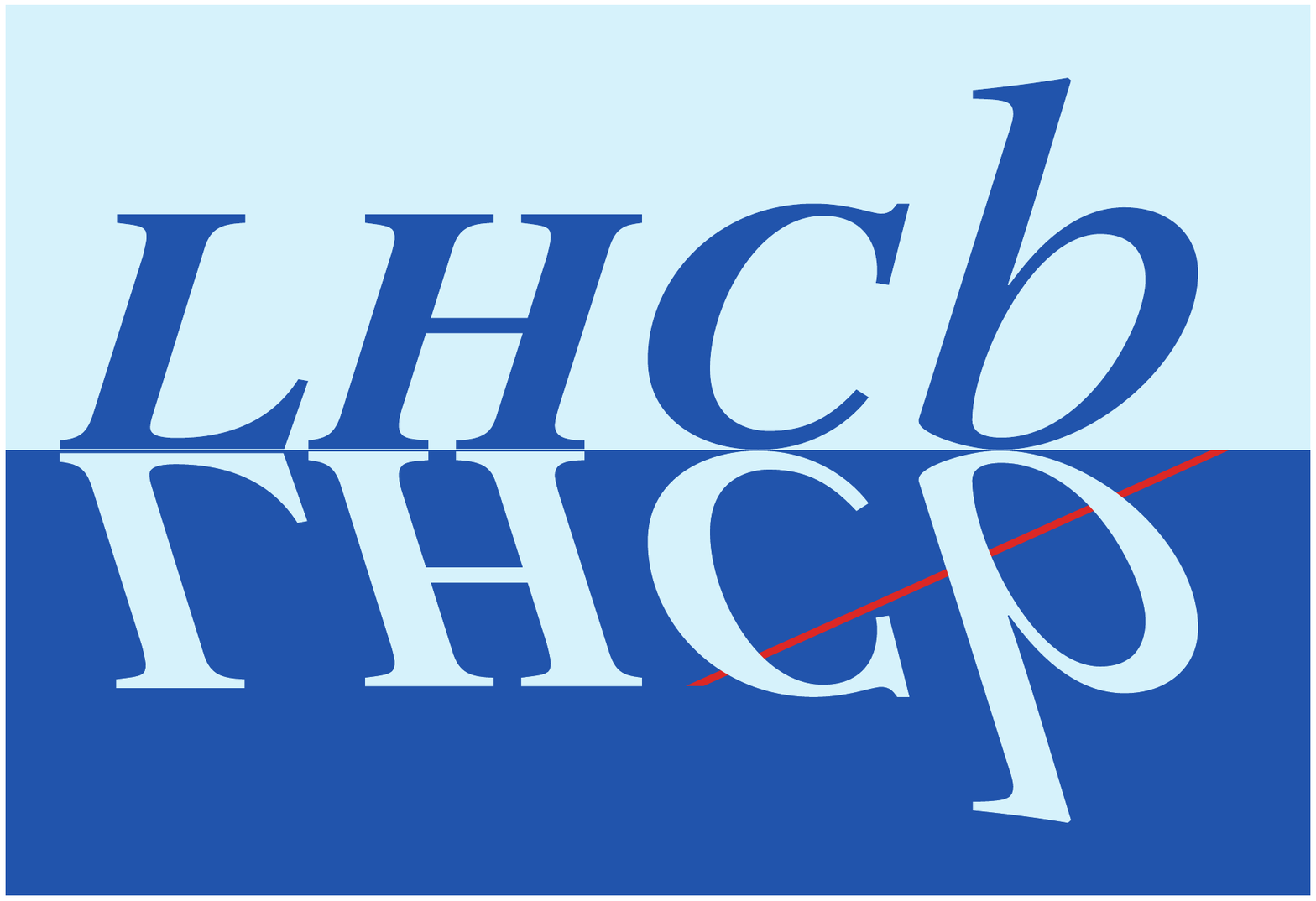}} & &}%
{\vspace*{-1.2cm}\mbox{\!\!\!\includegraphics[width=.12\textwidth]{lhcb-logo.eps}} & &}%
\\
 & & CERN-EP-2017-037 \\  
 & & LHCb-PAPER-2017-002\\  
 & & 14 March 2017 \\ 
 & & \\
\end{tabular*}

\vspace*{4.0cm}

{\normalfont\bfseries\boldmath\huge
\begin{center}
  Observation of five new narrow $\Omegac$ states decaying to $\Xicp \Km$ 
\end{center}
}

\vspace*{2.0cm}

\begin{center}
The LHCb collaboration\footnote{Authors are listed at the end of this paper.}
\end{center}

\vspace{\fill}

\begin{abstract}
  \noindent
The $\Xicp \Km$  mass spectrum is studied with a sample of \proton\proton collision data corresponding to an integrated luminosity of 3.3\invfb, collected by the LHCb experiment.
The $\Xicp$ is reconstructed in the decay mode \pkpi.
 Five new, narrow excited $\Omegac$ states are observed: the $\Omegares_\cquark(3000)^0$, $\Omegares_\cquark(3050)^0$, $\Omegares_\cquark(3066)^0$, $\Omegares_\cquark(3090)^0$, and $\Omegares_\cquark(3119)^0$. 
Measurements of their masses and widths are reported.  
\end{abstract}

\vspace*{2.0cm}

\begin{center}
  Published in Phys.~Rev.~Lett. 118 (2017) 182001 
\end{center}

\vspace{\fill}

{\footnotesize 
\centerline{\copyright~CERN on behalf of the \lhcb collaboration, licence \href{http://creativecommons.org/licenses/by/4.0/}{CC-BY-4.0}.}}
\vspace*{2mm}

\end{titlepage}


\newpage
\setcounter{page}{2}
\mbox{~}
%
%
%
%

\cleardoublepage

}

\renewcommand{\thefootnote}{\arabic{footnote}}
\setcounter{footnote}{0}



\pagestyle{plain} 
\setcounter{page}{1}
\pagenumbering{arabic}

The spectroscopy of singly charmed baryons  $\cquark \quark \quark^{\prime}$ is intricate. With three quarks and numerous degrees of freedom, many 
states are expected. At the same time, the large mass difference between the charm quark and the light quarks
provides a natural way to understand the spectrum by using the symmetries provided by Heavy Quark 
Effective Theory (HQET)~\cite{Grozin:1992yq,Mannel:1996rg}. 
In recent years, considerable improvements have been made in the predictions of the properties of these heavy baryons~\cite{Ebert:2007nw, Roberts:2007ni, Garcilazo:2007eh, Migura:2006ep, Ebert:2011kk, Valcarce:2008dr, Shah:2016nxi, Vijande:2012mk,Yoshida:2015tia,Chen:2015kpa,Chen:2016phw,Chiladze:1997ev}. 
In many of these models, the heavy quark interacts with a  $(\quark \quark^{\prime})$ diquark, which is treated as a single object. These models predict seven states in the mass range $2.9$--$3.2\gev$,\footnote{Natural units are used throughout the paper.} some of them narrow. Other models make use of Lattice QCD calculations~\cite{Padmanath:2013bla}.

The spectroscopy of charmed baryons, particularly the \Lc, {{\ensuremath{\Sigmares_\cquark}}\xspace}, and \Xic states, has also seen considerable experimental progress, with results obtained
at the $B$ factories and is in the physics program of the LHCb experiment at CERN~\cite{PDG2016,Aaij:2013voa}.
Among the expected charmed baryon states, this work addresses the $\Omegac$ baryons, which have quark content \cquark \squark \squark and isospin zero. Their spectrum is largely unknown:
only the $\Omegac$ and $\Omegares_\cquark(2770)^0$, presumed to be the $J^P=1/2^+$ and $3/2^+$ ground states, have been observed~\cite{PDG2016,Solovieva:2008fw}.

To improve the understanding of this little-explored sector of the charmed baryon spectrum, this Letter presents a search for new $\Omegac$ resonances that decay strongly to the final state $\Xicp \Km$, where the $\Xicp$ is a weakly decaying charmed baryon with quark content \cquark \squark \uquark.\footnote{The inclusion of charge-conjugate processes is implied throughout, unless stated otherwise.}
The measurement is based on samples of \proton\proton collision data corresponding to
integrated luminosities of $1.0$, $2.0$ and $0.3\invfb$ at center-of-mass energies
of 7, 8 and 13\tev, respectively, recorded by the LHCb experiment.
The LHCb detector
is a single-arm forward spectrometer covering the \mbox{pseudorapidity} range
$2<\eta <5$, designed for the study of particles containing \bquark or
\cquark quarks, and is described in detail in
Refs.~\cite{Alves:2008zz,LHCb-DP-2014-002}. Hadron identification is provided by two ring-imaging Cherenkov detectors~\cite{LHCb-DP-2012-003}, a calorimeter system, and a muon detector. 
The online event selection is
performed by a trigger, which consists of a hardware stage, based on information
from the calorimeter and muon systems, followed by a software stage, which
applies a full event reconstruction~\cite{LHCb-DP-2012-004}. Simulated events are produced with the software packages
described in Refs.~\cite{Sjostrand:2006za,LHCb-PROC-2010-056,Lange:2001uf,Golonka:2005pn,Agostinelli:2002hh,*Allison:2006ve,LHCb-PROC-2011-006}.

The reconstruction begins with the \Xicp baryon, via the decay $\Xicp \to \pkpi$. The \Xicp candidates are formed from combinations of 
three tracks that originate from a common vertex. These are required to pass a cut-based preselection and then a multivariate selection based on likelihood ratios, 
described below. Candidates fulfilling these requirements are then combined with a fourth track to form $\Omegac \to \Xicp \Km$ candidates to which additional selection requirements, also described below, are applied.

\begin{figure} [!ht]
\ifthenelse{\boolean{wordcount}}{}{
  \centering
  \includegraphics[width=0.56\textwidth]{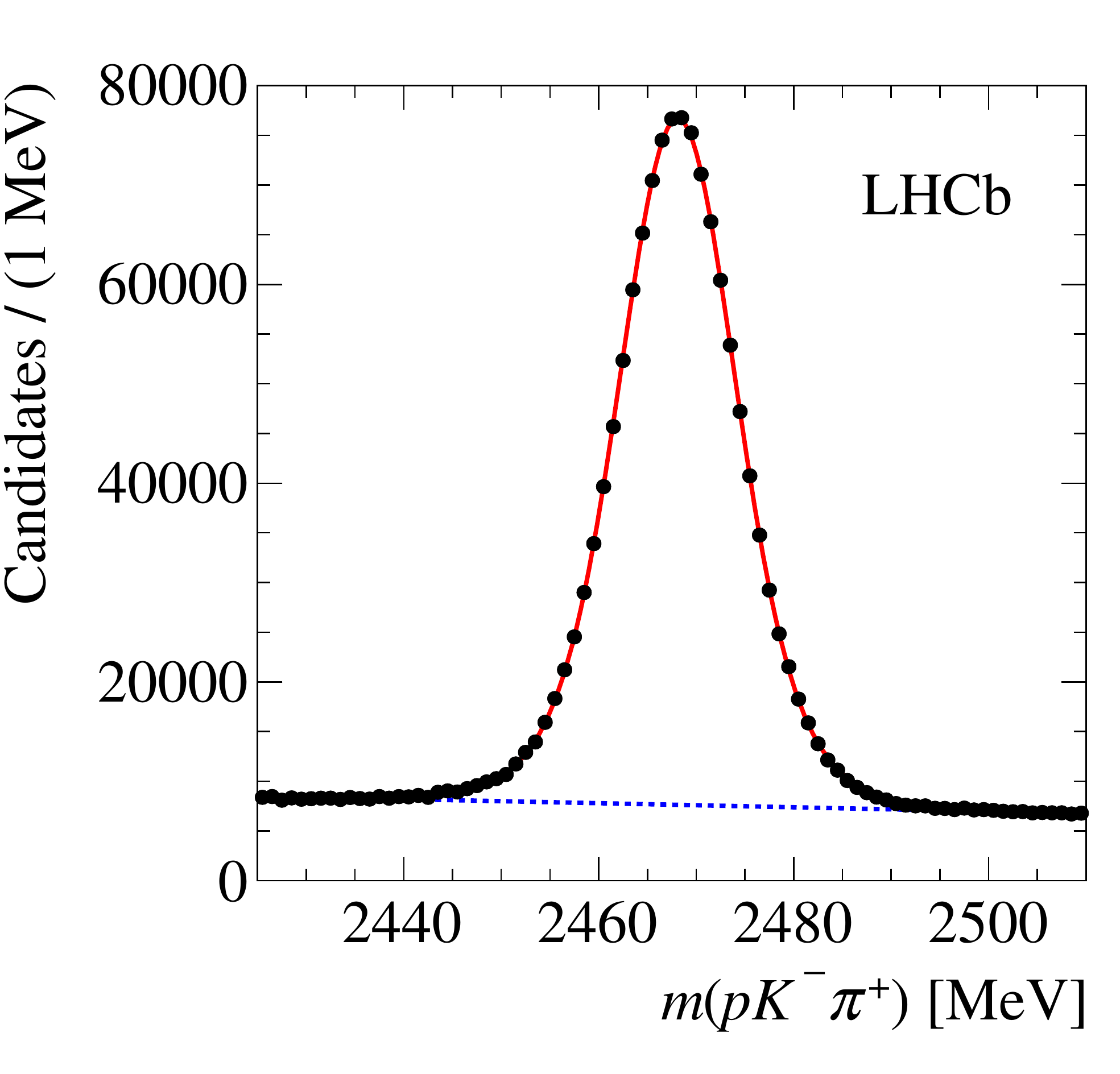}
}
  \caption{
Distribution of the reconstructed invariant mass $m(\pkpi)$ for all candidates in the inclusive \Xicp sample passing the likelihood ratio selection described in the text.
The solid (red) curve shows the result of the fit, and the dashed (blue) line indicates the fitted background.
  } 
  \label{fig:fig1}
\end{figure}

The \Xicp preselection requires a positively identified proton and a large \Xicp flight-distance significance (defined as the measured flight distance divided by its uncertainty) from a primary \proton\proton interaction vertex (PV). 
The \Xicp candidates are also constrained to originate from the PV by requiring a small \chisqip (defined as  the difference between the vertex-fit \chisq of the PV reconstructed with and without the candidate in question).
The resulting \pkpi mass spectrum is fitted with a linear function to describe the background and the sum of two Gaussian functions with a
common mean to describe the signal. 
The fit is used to define signal and sideband regions of the \Xicp invariant mass spectrum: the signal region consists of the range within $\pm 2.0\, \sigma$ of the fitted mass, where $\sigma = 6.8\mev$ is the weighted average of the standard deviations of the Gaussian functions, and the sidebands cover the range $3.5$--$5.5\, \sigma$ on either side.
The fit is also used to determine the \Xicp purity after the preselection, defined as the signal yield in the signal region divided by the total yield in the same region. A purity of 41\% is obtained, which is not sufficient for the spectroscopy study, 
but allows the extraction of background-subtracted probability density functions (PDFs) of the kinematic and geometric 
properties of the signal. These distributions are taken from data rather than simulation, 
given the limited understanding of heavy baryon production dynamics and the difficulty of modeling 
them correctly for different center-of-mass energies.

For each variable of interest the background PDF is obtained from the corresponding distribution in the mass sideband regions, and is also used for the background subtraction.
The signal \PDF is obtained from the normalized, background-subtracted distribution in the signal mass region. 
Variables found to have a good discrimination between signal and background are: the vertex fit \chisq, the 
\Xicp flight-distance significance and \chisqip, the particle identification probability for the proton and the kaon from the \Xicp decay, 
the \chisqip of the three individual tracks,
the \Xicp transverse momentum \pt with respect to the
beam axis, the pseudorapidity $\eta$, and the angle between the \Xicp momentum and the vector joining the
PV and the \Xicp decay vertex.
 
The PDFs of the 11 variables (${\bf x}$) above are used to form a likelihood ratio, whose logarithm is defined as
\begin{equation}
\calL(\mathbf{x}) = \sum_{i=1}^{11}[\ln{\rm \PDF_{sig}}({\bf\it x}_{i}) - \ln{\rm \PDF_{back}}({\bf\it x}_{i})],
\end{equation}
where ${\rm \PDF_{sig}}$ and ${\rm \PDF_{back}}$ are the \PDF distributions for signal and background, respectively. 
Correlations between the variables are neglected in the likelihood.

The likelihood ratios and their PDFs are defined separately for the three data sets at different center-of-mass energies due to their different
trigger conditions. 
The selection requirements on the likelihood ratios are also chosen separately for the three samples, and lead to 
\Xicp purities of approximately 83\% in the inclusive \Xicp sample.

Figure~\ref{fig:fig1} shows the \pkpi mass spectrum of \Xicp candidates passing the likelihood ratio selection 
for all three data sets combined, along with the result of a fit with the functional form described above.
The \Xicp signal region contains $1.05 \times 10^6$ events.
Note that this inclusive \Xicp sample contains not only those produced in the decays of charmed baryon resonances but also from other sources, including decays of \bquark hadrons and direct production at the PV.

Each \Xicp candidate passing the likelihood ratio selection and lying within the \Xicp signal mass region is then combined in turn with each \Km candidate in the event. 
A vertex fit is used to reconstruct each \Xicp \Km combination, with the constraint that it originates from the PV.
The $\Xicp \Km$ candidate must have a small vertex fit \chisq, a high kaon identification probability,
and transverse momentum $\pt(\Xicp \Km)>4.5\gev$.

\begin{figure} [tbp]
\ifthenelse{\boolean{wordcount}}{}{
  \centering
  \includegraphics[width=0.67\textwidth]{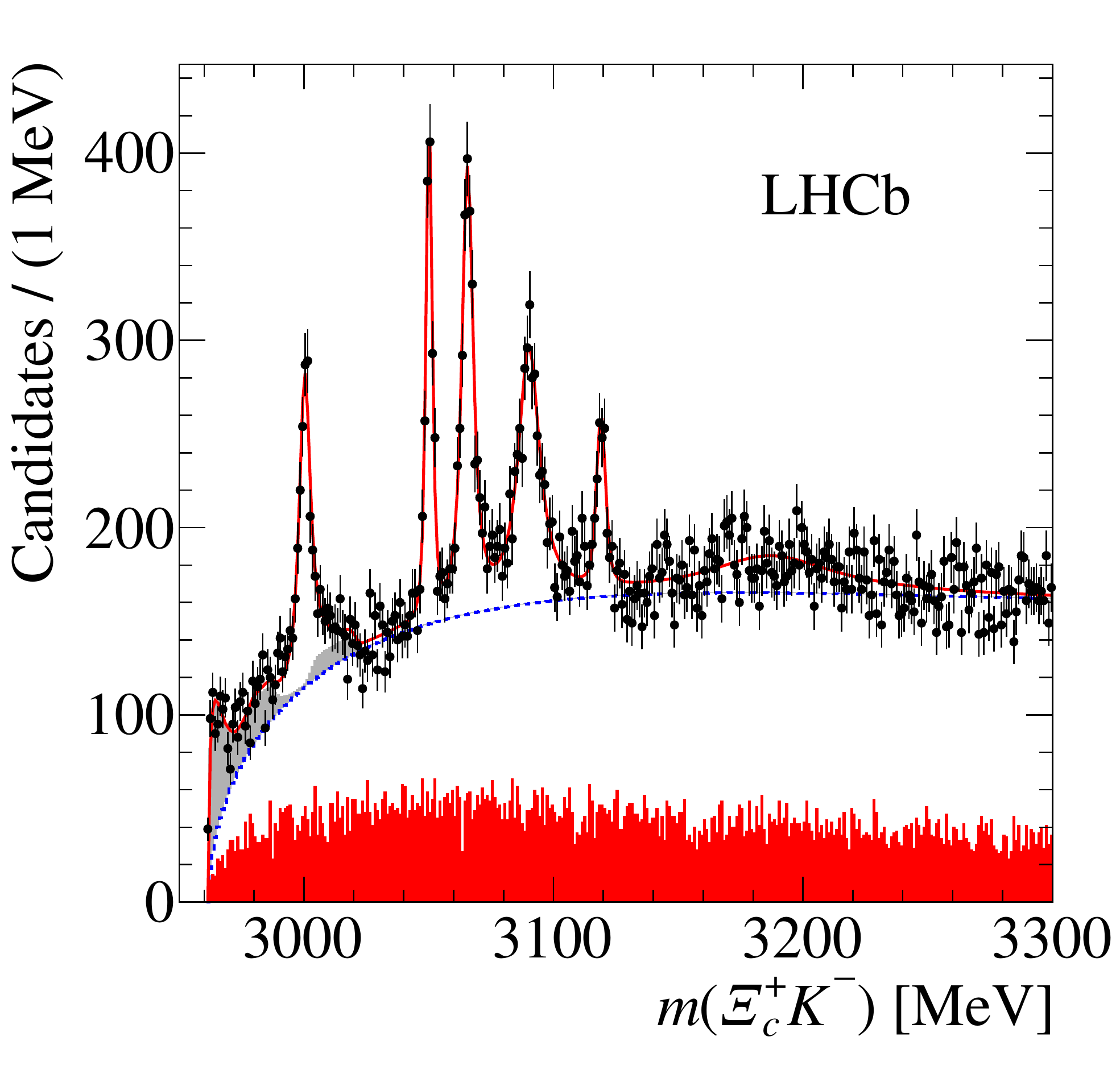}
}
  \caption{Distribution of the reconstructed invariant mass $m(\Xicp \Km)$ for all candidates passing the likelihood ratio selection; the solid (red) curve
shows the result of the fit, and the dashed (blue) line indicates the fitted background.
The shaded (red) histogram shows the corresponding mass spectrum from the \Xicp sidebands and the shaded (light gray) distributions
indicate the feed-down from partially reconstructed $\Omegares_\cquark(X)^0$ resonances. 
  } 
  \label{fig:fig2}
\end{figure}

\begin{figure} [!htb]
\ifthenelse{\boolean{wordcount}}{}{
  \centering
  \includegraphics[width=0.67\textwidth]{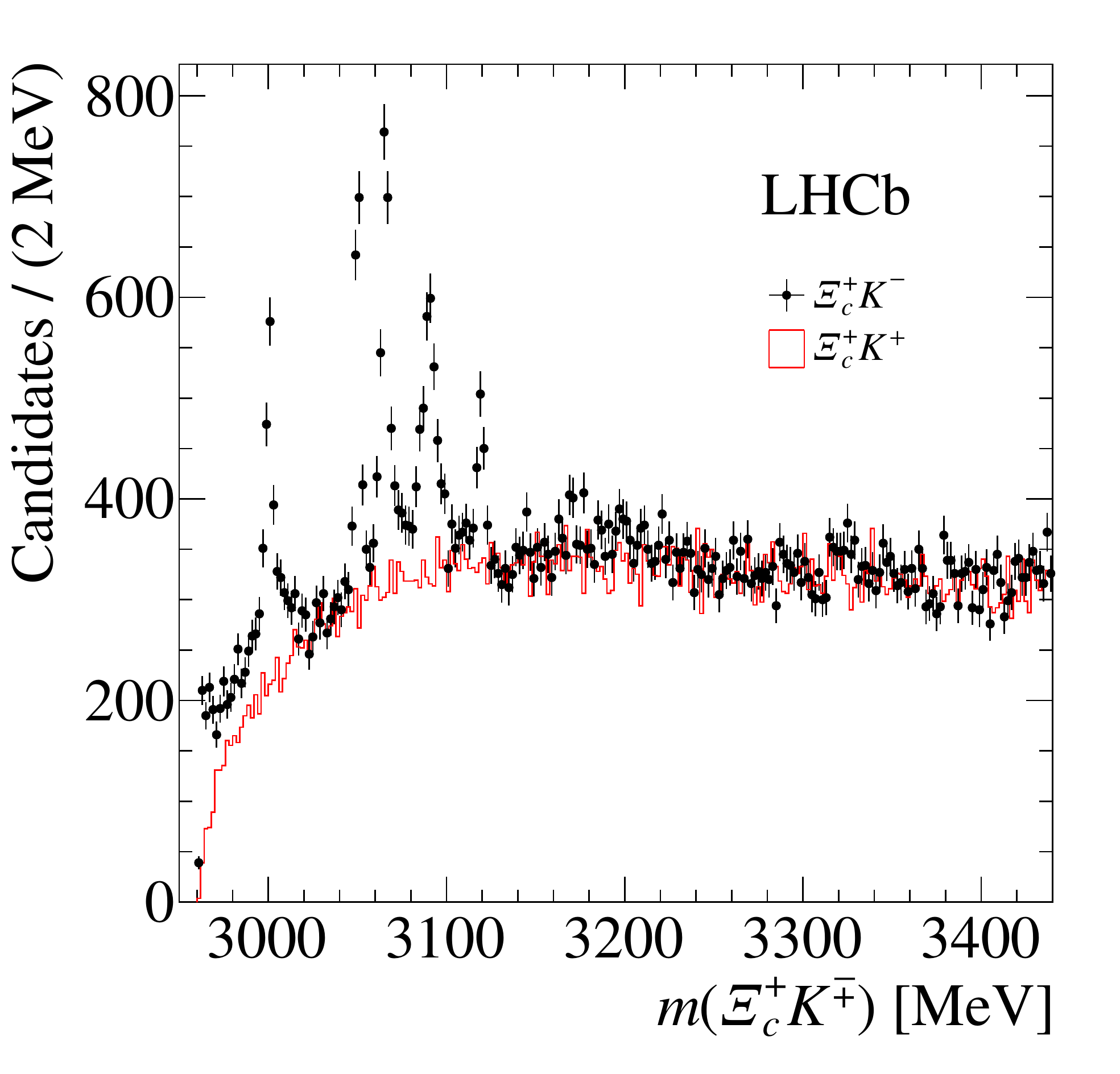}
}
  \caption{
Distribution of the reconstructed invariant mass $m(\Xicp \Km)$ for all candidates passing the likelihood ratio selection, shown as black points with error bars, and the wrong-sign $m(\Xicp \Kp)$ spectrum scaled by a factor of 0.95, shown as a solid (red) histogram. 
  } 
  \label{fig:fig3}
\end{figure} 

The $\Xicp \Km$ invariant mass is computed as
\begin{equation}
m(\Xicp \Km) = m([\pkpi]_\Xicp \Km) - m([\pkpi]_\Xicp) + m_\Xicp,
\end{equation}
where $m_\Xicp=~2467.89^{+0.34}_{-0.50}~\mev$ is the world-average \Xicp mass~\cite{PDG2016} and $[\pkpi]_\Xicp$ is the reconstructed $\Xicp \to \pkpi$ candidate.

 In this analysis, the distribution of the invariant mass $m(\Xicp \Km)$  is studied from threshold up to 3450\mev.  

The $\Xicp \Km$ mass distribution for the combined data sets is shown in Fig.~\ref{fig:fig2} where five narrow structures are observed.
To investigate the origin of these structures, Fig.~\ref{fig:fig2} also shows the distribution of $m(\Xicp \Km)$ in the \Xicp 
sidebands as a shaded (red) histogram; no structure is seen in this background sample.
In addition, wrong-sign $\Xicp \Kp$ combinations are processed in the same way as the right-sign combinations. The resulting wrong-sign $\Xicp \Kp$ mass spectrum is shown in Fig.~\ref{fig:fig3}, 
scaled by a factor of 0.95 so that the two spectra approximately match at large invariant mass,
along with the right-sign $m(\Xicp \Km)$ spectrum for comparison.
No structure is observed in the wrong-sign mass spectrum.
The absence of corresponding features in the control samples is consistent with the five structures being resonant states, 
henceforth denoted $\Omegares_\cquark(X)^0$ for mass $X$.
It can also be seen that the two mass spectra in Fig.~\ref{fig:fig3} exhibit different behavior close to the
$\Xicp \Km$ threshold (2960--2970\mev). 
The right-sign distribution has a much steeper rise 
compared to the wrong-sign spectrum, suggesting the presence of additional components in the $\Xicp \Km$ mass spectrum as discussed below.

Further tests are performed by
studying combinations of one of the $\Xicp \to \pkpi$ decay products with the other kaon used to form the $\Omegares_\cquark(X)^0$ candidate 
(\ie $\proton \Km$, $\Km \Km$, $\pip \Km$).
The resulting two-body invariant mass spectra 
do not show any structure except for a small \Kstarzb signal in the $\pip \Km$ mass, also visible in the \Xicp sidebands,  which is attributed to background contributions.
Another class of potential misreconstruction consists of $\Xicp \pim$ combinations in which the $\pim$ is misidentified as a kaon. 
To test for this, the selected $\Xicp \Km$ sample is investigated with the pion mass assigned to the kaon candidate. 
No narrow peaks are observed in this pseudo-$\Xicp \pim$ spectrum, 
indicating that peaks in the $\Xicp \Km$ spectrum do not arise from misidentified $\Xicp \pim$ resonances.

The wrong-sign $\Xicp \Kp$ sample is used to study the combinatorial background.
The parameterization used is~\cite{delAmoSanchez:2010vq} 
\begin{equation}
B(m) = \left\{
\begin{array}{l}
 P(m)e^{a_1m+a_2m^2} \ \ \ \ {\rm for} \ m<m_0, \\
 P(m)e^{b_0+b_1m+b_2m^2} \ {\rm for} \ m>m_0,
\end{array}
\right.
  \label{eq:back}
\end{equation}
where $P(m)$ is a two-body phase-space factor and $m_0$, $a_i$ and $b_i$ are free parameters. 
Both $B(m)$ and its first derivative must be continuous at $m=m_0$; these constraints reduce the number of free parameters to four.
This model gives a good description of the wrong-sign mass spectrum up to a mass of 3450\mev with a $p$-value of 18\% for a binned \chisq fit.

To study the reconstruction efficiency and the mass resolution of each of the structures, samples of simulated events are generated in which
$\Omegares_\cquark(X)^0$ resonances decay to $\Xicp \Km$, with the masses and natural widths of the $\Omegares_\cquark(X)^0$ 
chosen to approximately match those seen in data.
The mass residuals, defined as the difference between the generated $\Omegares_\cquark(X)^0$ mass and the reconstructed value of $m(\Xicp \Km)$,
are well described by the sum of two Gaussian functions with a common mean. The parameters of
these fits are used to determine
the mass-dependent experimental resolution, which runs from 0.75\mev at 3000\mev  to 1.74\mev at 3119\mev,
and is found to be well described by a linear function.
The simulation samples are also used to obtain the reconstruction efficiency, which is consistent with being constant as a function of 
$m(\Xicp \Km)$.

Another possible decay mode for $\Omegares_\cquark(X)^0$ resonances is 
\begin{equation}
\Omegares_\cquark(X)^0 \to \Km \PXi_c^{\prime+ } \quad\mbox{with}\quad\PXi_c^{\prime+ } \to \Xicp \gamma,
\label{eq:cf}
\end{equation}
or, in general, to a final state that includes $\Xicp \Km$ but also contains one or more additional particles that are not included in the reconstruction. For the case of a narrow $\Omegares_\cquark(X)^0$ resonance decaying via $\PXi_c^{\prime+}$, the resulting distribution in $m(\Xicp \Km)$
is a relatively narrow structure that is shifted down in mass (feed-down)  that needs to be taken into account
in the description of the data.
Simulation studies of the decay chain shown in Eq.~\ref{eq:cf}
have been performed with resonance masses of 3066, 3090, and 3119\mev.
It is found that the feed-down shapes deviate from Breit-Wigner distributions and are therefore parameterized by B-splines~\cite{DeBoor1428148}.

A binned $\chi^2$ fit to the $m(\Xicp \Km)$ spectrum is performed in the range from threshold to 3450\mev.
In this fit, the background is modeled by Eq.~\ref{eq:back},
while the resonances are described by spin-zero relativistic Breit--Wigner functions convolved with the experimental resolution. 
In addition, three feed-down contributions arising from the partially reconstructed decays of $\Omegares_\cquark(3066)^0$, $\Omegares_\cquark(3090)^0$, and $\Omegares_\cquark(3119)^0$ resonances are included with fixed shapes but free yields.
It is found that the fit improves if an additional broad Breit--Wigner function is included in the 3188\mev mass region.
This broad structure may be due to a single resonance, to the superposition of several resonances, to feed-down from higher states, or to some combination of the above. Under the simplest hypothesis, namely that it is due to a single state, its parameters are given in Table~\ref{tab:tab1}.

This configuration is denoted the reference fit, and is shown in Fig.~\ref{fig:fig2}. No significant structure is seen above 3300\mev.
Table~\ref{tab:tab1} gives the fitted parameters and yields of the resonances, along with the yields
for the feed-down contributions indicated with the  subscript ``fd''.  
The statistical significance of each resonance is computed as $N_{\sigma}=\sqrt{\Delta \chisq}$, where $\Delta \chisq$ is the increase in \chisq when the resonance is excluded in the fit.
Very high significances are obtained for all the narrow resonances observed in the mass spectrum.
The threshold enhancement below 2970\mev is fully explained by feed-down from the $\Omegares_\cquark(3066)^0$ resonance.
 \begin{table*}[bt]
\centering
  \caption{Results of the fit to $m(\Xicp \Km)$ for the mass, width, yield and significance for each resonance. 
The subscript ``${\rm fd}$'' indicates the feed-down contributions described in the text.
For each fitted parameter, the first uncertainty is statistical and the second systematic.
The asymmetric uncertainty on the $\Omegares_\cquark(X)^0$ arising from the \Xicp mass is given separately.
Upper limits are also given for the resonances $\Omegares_\cquark(3050)^0$ and $\Omegares_\cquark(3119)^0$ for which the width is not significant.
}
\ifthenelse{\boolean{wordcount}}{}{
\resizebox{\textwidth}{!}{
  \begin{tabular}{lcccc} 
    \hline
\noalign{\vskip2pt}
Resonance  &  Mass (\mev) & $\Gamma$ (\mev) &  Yield & $N_{\sigma}$\cr
\hline
\noalign{\vskip2pt}
$\Omegares_\cquark(3000)^0$ & $3000.4 \pm 0.2 \pm 0.1^{+0.3}_{-0.5}$ & $4.5 \pm 0.6 \pm 0.3$ & $1300 \pm 100 \pm \al80$ & 20.4\cr
\noalign{\vskip2pt}
$\Omegares_\cquark(3050)^0$ & $3050.2 \pm 0.1 \pm 0.1 ^{+0.3}_{-0.5}$ & $0.8 \pm 0.2 \pm 0.1$ & $\al 970 \pm \al60 \pm \al20$ & 20.4 \cr
\noalign{\vskip2pt}
                     &                                    & $< 1.2 \mev, 95\%$ {\rm CL} & \cr
\noalign{\vskip2pt}
$\Omegares_\cquark(3066)^0$ & $3065.6 \pm 0.1 \pm 0.3 ^{+0.3}_{-0.5}$ & $3.5 \pm 0.4 \pm 0.2$ & $1740 \pm 100 \pm \al50$ &  23.9\cr
\noalign{\vskip2pt}
$\Omegares_\cquark(3090)^0$ & $3090.2 \pm 0.3 \pm 0.5 ^{+0.3}_{-0.5}$ & $8.7 \pm 1.0\pm 0.8 $ & $ 2000 \pm 140\pm 130$ & 21.1 \cr
\noalign{\vskip2pt}
$\Omegares_\cquark(3119)^0$ & $3119.1 \pm 0.3 \pm 0.9 ^{+0.3}_{-0.5}$ & $1.1 \pm 0.8\pm 0.4 $ & $\al 480 \pm \al 70\pm \al 30$ & 10.4 \cr
\noalign{\vskip2pt}
                     &                                    & $< 2.6 \mev, 95\%$ {\rm CL} & \cr
\noalign{\vskip2pt}
\hline
\noalign{\vskip2pt}
$\Omegares_\cquark(3188)^0$ & $\phantom{1.}3188 \pm \phantom{0}5\phantom{.} \pm 13\phantom{.^{+0.3}_{-0.5}}$ & $60 \pm \all 15 \pm 11 $& $1670 \pm 450 \pm 360$ & \cr
\noalign{\vskip2pt}
\hline
\noalign{\vskip2pt}
$\Omegares_\cquark(3066)^0_{\rm fd}$ &  &  & $\al 700 \pm \al 40 \pm 140$ &   \cr
\noalign{\vskip2pt}
$\Omegares_\cquark(3090)^0_{\rm fd}$ &  &  & $\al 220 \pm \al 60 \pm \al 90 $ &  \cr
\noalign{\vskip2pt}
$\Omegares_\cquark(3119)^0_{\rm fd}$ &  &  & $\al 190 \pm \al 70 \pm \al 20$ &   \cr
\hline
  \end{tabular}
}
}
  \label{tab:tab1}
 \end{table*}

Several additional checks are performed to verify the presence of the signals and the stability of the fitted parameters.
The likelihood ratio requirements are varied, testing both looser and tighter selections.
As another test, the data are divided into subsamples according to the data-taking conditions, and each subsample is analyzed and fitted separately.
The charge combinations $\Xicbarm \Kp$ and $\Xicp \Km$ are also studied separately.
 In all cases  the fitted resonance parameters
are consistent among the subsamples and with the results from the reference fit.

Systematic uncertainties on the \Omegac resonance parameters are evaluated as follows.
The fit bias is evaluated by generating and fitting an ensemble of 500 random mass spectra that are generated according to
the reference fit. 
For each parameter, the absolute value of the difference between the input value and the mean fitted value of the ensemble is taken as the systematic uncertainty.

The background model uncertainty is estimated by exchanging it for the alternative function
$B^{\prime}(m) = (m - m_{\rm th})^\alpha e^{\beta + \gamma m + \delta m^2}$, where $m_{\rm th}$ is the threshold mass and $\alpha, \beta, \gamma $ and $\delta$ are free parameters.
The uncertainty associated with the choice of the Breit--Wigner model is estimated
by fitting the data with relativistic $L=1,2$ Breit--Wigner functions with varying Blatt--Weisskopf factors~\cite{Blatt}, and is found to be negligible.

Resonances can interfere if they are close in mass and have the same spin-parity.
The effect is studied by introducing interference terms between each resonance and its neighboring resonances, one pair of resonances at a time.
This is implemented with an amplitude of the form $A = |c_{i}{\rm BW}_{i} + c_j {\rm BW}_{j}e^{i \phi}|^2$ for the interference between resonances $i$ and $j$, 
where ${\rm BW}_{i}$ and ${\rm BW}_{j}$ are complex Breit--Wigner functions and $c_{i,j}$ and $\phi$ are free parameters. 
For the central three resonances where interference could occur with the state to the left or to the right,
the absolute values of the deviations are added in quadrature. No evidence for interference effects is observed.

Recently, the Belle collaboration has reported a measurement of the $\PXi_\cquark^{\prime+}$ mass~\cite{Yelton:2016fqw} that is significantly more precise than the previous value, and which differs from it by +2.8\mev. The effect of this is tested by shifting the $\Omegares_\cquark(3066)^0$, 
$\Omegares_\cquark(3090)^0$, and $\Omegares_\cquark(3119)^0$ feed-down shapes accordingly, and it is included as a systematic uncertainty.

The mass scale uncertainty is studied with a series of control samples and is found to be 0.03\% of the mass difference from
threshold $(m - m_{\rm th})$.
A comparison between the fitted \Xicp mass resolution in data and simulation shows a 1.7\% discrepancy, which is assigned as a systematic uncertainty 
on the width of the resonances.
The description of the broad, high-mass structure labeled $\Omegares_\cquark(3188)^0$ is changed to the sum of four incoherent Breit--Wigner functions and the 
effect on the other five resonances is included in the list of the systematic uncertainties. The largest contribution is found to be from possible interference, while the feed-down shift has a sizeable effect only on the $\Omegares_\cquark(3000)^0$ parameters.
For the total systematic uncertainty the individual contributions are added in quadrature.
Finally, an uncertainty arises from the uncertainty on the $\Xicp$ mass, whose world-average value is $m(\Xicp) = 2467.89^{+0.34}_{-0.50}\mev$~\cite{PDG2016}. It is quoted separately from the other uncertainties on the resonance masses, and is the dominant uncertainty on several of them.

The $\Omegares_\cquark(3050)^0$ and $\Omegares_\cquark(3119)^0$ resonances have very narrow widths.
For these states, Table~\ref{tab:tab1} also includes Bayesian 95\% confidence level (CL) upper limits~\cite{PDG2016} on the widths, evaluated from
the statistical and systematic uncertainties assuming Gaussian PDFs.

The observation of these $\Omegares_\cquark$ states in an inclusive process through a two-body decay does not allow the determination of their quantum numbers,
and therefore no attempt is made to compare the measured masses with HQET expectations. 
More information can be obtained from the study of possible three-body decays or when reconstructing these states in
decays of heavy baryons.

In conclusion, the $\Xicp \Km$ mass spectrum is investigated using a data set corresponding to an integrated luminosity of 3.3\invfb
collected by the LHCb experiment. 
A large and high-purity sample of \Xicp baryons is reconstructed in the Cabibbo-suppressed decay mode \pkpi.
Five new, narrow excited $\Omegac$ states are observed: the $\Omegares_\cquark(3000)^0$, $\Omegares_\cquark(3050)^0$, $\Omegares_\cquark(3066)^0$, $\Omegares_\cquark(3090)^0$, and $\Omegares_\cquark(3119)^0$, and measurements of their masses and widths are reported. 
The data indicate also the presence of a broad structure around 3188\mev that is
fitted as a single resonance but could be
produced in other ways, for example as a superposition of several states. 
In addition, the partially reconstructed decay $\Omegares_\cquark(3066)^0 \to \PXi_\cquark^{\prime+} \Km$ is observed via its feed-down in the threshold region.
Similarly, indications are found of $\Omegares_\cquark(3090)^0$ and $\Omegares_\cquark(3119)^0$  decays to $\PXi_\cquark^{\prime+} \Km$.

\ifthenelse{\boolean{wordcount}}{}{
\section*{Acknowledgements}

\noindent We express our gratitude to our colleagues in the CERN
accelerator departments for the excellent performance of the LHC. We
thank the technical and administrative staff at the LHCb
institutes. We acknowledge support from CERN and from the national
agencies: CAPES, CNPq, FAPERJ and FINEP (Brazil); MOST and NSFC (China);
CNRS/IN2P3 (France); BMBF, DFG and MPG (Germany); INFN (Italy); 
NWO (The Netherlands); MNiSW and NCN (Poland); MEN/IFA (Romania); 
MinES and FASO (Russia); MinECo (Spain); SNSF and SER (Switzerland); 
NASU (Ukraine); STFC (United Kingdom); NSF (USA).
We acknowledge the computing resources that are provided by CERN, IN2P3 (France), KIT and DESY (Germany), INFN (Italy), SURF (The Netherlands), PIC (Spain), GridPP (United Kingdom), RRCKI and Yandex LLC (Russia), CSCS (Switzerland), IFIN-HH (Romania), CBPF (Brazil), PL-GRID (Poland) and OSC (USA). We are indebted to the communities behind the multiple open 
source software packages on which we depend.
Individual groups or members have received support from AvH Foundation (Germany),
EPLANET, Marie Sk\l{}odowska-Curie Actions and ERC (European Union), 
Conseil G\'{e}n\'{e}ral de Haute-Savoie, Labex ENIGMASS and OCEVU, 
R\'{e}gion Auvergne (France), RFBR and Yandex LLC (Russia), GVA, XuntaGal and GENCAT (Spain), Herchel Smith Fund, The Royal Society, Royal Commission for the Exhibition of 1851 and the Leverhulme Trust (United Kingdom).

\addcontentsline{toc}{section}{References}
\setboolean{inbibliography}{true}
\bibliographystyle{LHCb}
\bibliography{main,LHCb-PAPER,LHCb-DP}}
 \setboolean{inbibliography}{false}

\newpage
\ifthenelse{\boolean{wordcount}}{}{
 
\clearpage
\centerline{\large\bf LHCb collaboration}
\begin{flushleft}
\small
R.~Aaij$^{40}$,
B.~Adeva$^{39}$,
M.~Adinolfi$^{48}$,
Z.~Ajaltouni$^{5}$,
S.~Akar$^{59}$,
J.~Albrecht$^{10}$,
F.~Alessio$^{40}$,
M.~Alexander$^{53}$,
S.~Ali$^{43}$,
G.~Alkhazov$^{31}$
% ...

{\footnotesize \it
% ...
$ ^{5}$Clermont Universit{\'e}, Universit{\'e} Blaise Pascal, CNRS/IN2P3, LPC, Clermont-Ferrand, France\\
% ...
$ ^{10}$Fakult{\"a}t Physik, Technische Universit{\"a}t Dortmund, Dortmund, Germany\\
% ...
$ ^{31}$Petersburg Nuclear Physics Institute (PNPI), Gatchina, Russia\\
% ...
$ ^{39}$Universidad de Santiago de Compostela, Santiago de Compostela, Spain\\
$ ^{40}$European Organization for Nuclear Research (CERN), Geneva, Switzerland\\
% ...
$ ^{43}$Nikhef National Institute for Subatomic Physics, Amsterdam, The Netherlands\\
% ...
$ ^{48}$H.H. Wills Physics Laboratory, University of Bristol, Bristol, United Kingdom\\
% ...
$ ^{53}$School of Physics and Astronomy, University of Glasgow, Glasgow, United Kingdom\\
% ...
$ ^{59}$University of Cincinnati, Cincinnati, OH, United States\\
% ...
}
\end{flushleft}
}
\end{document}